\documentclass[twocolumn,showpacs,preprintnumbers,amsmath,amssymb]{revtex4}
\usepackage{graphicx}% Include figure files
\usepackage{dcolumn}% Align table columns on decimal point
\usepackage{bm}% bold math
\begin{document}

\title{On Open Inflation, the string theory landscape and the low CMB
quadrupole}

\author{Michael Barnard}
 \email{barnard@physics.ucdavis.edu}
 \author{Andreas Albrecht}
 \email{albrecht@physics.ucdavis.edu}
 \affiliation{Department of Physics \\ 
One Shields Avenue\\
University of California\\
Davis, CA 95616}

%\author{Second Author}
%\affiliation{}

\date{\today}

\begin{abstract}
Cosmologists have embraced a particular ad hoc formula for
the primordial power spectrum from inflation for universes with
$\Omega_0 < 1$.   However, the so-called ``Open Inflation'' models,
which are attracting renewed  interest in the context of the ``string 
theory landscape'' give a different result, and offer a more fully
developed picture of the cosmology and fundamental physics  basis for
inflation with $\Omega_0 < 1$. The Open Inflation power spectrum
depends not only on $\Omega_0$, but on the parameters of the effective
fields that drive the universe {\em before} the Big Bang (in ``another
part of the landscape'').  This paper 
considers the search for features in CMB temperature anisotropy data
that might reflect a primordial spectrum of the Open Inflation form.
We ask whether this search could teach us about high energy physics that
described the universe before the onset of the Big Bang, and perhaps
even account for the low CMB quadrupole. Unfortunately our
conclusion is that the specific features we consider are unobservable
even with future experiments although we note a possible loophole
connected with our use of the thin wall approximation.
\end{abstract}

\pacs{Valid PACS appear here}


\maketitle

{\em 
{\bf Note Added}: Since this paper was completed we became aware of a large
body of existing literature which treats the problem of perturbations
in Open Universe models with a much greater degree of sophistication
than we do here (including working away from the thin wall limit).
See \cite{Garriga:1996pg,Sasaki:1996qn,Garcia-Bellido:1996gd,Sasaki:1997ex,Garcia-Bellido:1997hy,Garriga:1997wz,Garcia-Bellido:1997te,Garriga:1998he,Garcia-Bellido:1998wd,Linde:1999wv} and references therein. 

An up-to-date 
treatment of the important questions raised in this paper (about the
possible universality of open inflation in the string theory landscaped
and resulting observational signatures) requires the application of
these more sophisticated methods and results, a process we are now
undertaking. We apologies to the authors of this impressive earlier
work for our ignorance about it in the first version of this paper
posted on the archive. We also thank Jaume Garriga and Thomas Hertog
for bringing this work to our attention. 

}

\section*{Introduction}

One of the great achievements of modern cosmology is the ability to
calculate detailed predictions for the cosmological perturbations from
specific models of the early universe. This, along with impressive new
data such as the WMAP survey\cite{Bennett:2003bz} has allowed
significant constraints to be placed on early universe physics as well
as on a number of cosmological parameters. 

One of the key cosmological parameters is $\Omega_0$, the ratio of the
current cosmic density (including the dark energy) to the critical
density.  A well-known problem is that for cosmological models with
$\Omega_0 < 1$ the perturbation calculation is more problematic,
particularly on large scales.  This is because 
for typical models of cosmic inflation to make precise predictions
for perturbations on all observed scales they must also predict
$\Omega_0 = 1$ to about one part in $10^5$.  In the context of these
models, to calculate the large scale perturbations in the $\Omega_0 <
1$ case one must answer the question ``what physics other than
inflation determined the perturbations on the largest observable scales?''. This issue has
been recognized since the first papers on inflation with
$\Omega_0 < 1$\cite{Lyth:1990dh,Ratra:1994dm}.

For the most part, the cosmology community has ``resolved'' this problem by
simply assuming a particular formula for perturbations in cosmologies
with $\Omega_0 < 1$. 
This formula appears in all the main software packages (such as
CMBfast) which determine the perturbation spectra for $\Omega_0 < 1$
models.  It 
is only because of this particular choice that it even seems possible
to determine $\Omega_0$ to high precision.  One is left open to the
possibility that a deeper understanding of early universe physics
could shift our preference to different 
pictures of $\Omega_0 < 1$ cosmology which could yield  different
formulas for the perturbation spectrum.  For $\Omega_0 < 1$ models with
different spectra, the same data might well lead to a different
preferred value of $\Omega_0$ as well as other parameters. 

In fact, we may be in the midst of such a shift right now.  Recent
work \cite{Kachru:2003aw} suggests that string theory (our best hope for a
realistic quantum gravity theory) predicts a landscape of different
``vacua'' which are highly stable, but which have some non-zero
probability of tunneling into one another.  This picture suggests a
cosmology strikingly similar to the so-called ``Open Inflation'' models of
Bucher et al. \cite{BT1,BT2}) 

The Open Inflation models were first invented to address the
ambiguities of the perturbation spectra for $ \Omega_0 < 1$
cosmologies discussed above.  Bucher et al. consider a
cosmological model with an initial phase of inflation that defines
the cosmological state on a range of length scales that spans many
orders of magnitude and drives the global state of the universe toward
$\Omega_0 = 1$. Bucher et al. modeled this phase of inflation with a
field trapped in false vacuum, in the manner of ``old
inflation''\cite{Guth:1980zm}.  

This initial period of inflation ends with a
tunneling process that produces a bubble universe which is open from
the point of view of observers within it.  The field that tunnels can
experience a shorter period of slow-roll inflation\cite{new} after the
tunneling event which can bring the bubble universe close to $\Omega =
1$ and define the perturbation spectrum on smaller scales.  Because of
the early period of old inflation the pre-tunneling cosmic state is 
uniquely determined, and this allows the perturbations in the bubble
universe to be well determined on {\em all} observable scales with no
ambiguities.  

When first introduced the open inflation models seemed a bit
artificial (although it really was a matter of taste whether one
considered them more so than ``typical'' slow-roll inflation models).
Today, the landscape picture that is emerging from string theory
suggests that the cosmology for a universe in any one of the
many metastable vacua universally starts with a tunneling event
preceded by a long period of old inflation in the (false) vacuum of the
previous landscape location. Although there still are a number of
unresolved questions, this picture certainly suggests that the Open
Inflation model of Bucher et al. may well be {\em the} universal cosmology seen
by an observer in the string theory landscape.\footnote{
Others have considered inflation in the string theory landscape (see
for example
\cite{Kachru:2003sx,Dvali:2003vv,Hsu:2003cy,Buchel:2003qj,Firouzjahi:2003zy,Pilo:2004mg,Linde:2004kg,Garousi:2004uf,Burgess:2004kv,Berg:2004ek,Buchel:2004qg,Dasgupta:2004dw}).
Linde gives a nice review in sections 11 and 12 of
\cite{Linde:2004kg}.  For 
the most part these authors have sought a sufficiently large
amount of new inflation at the onset of the big bang  to make the
issues raised here  unimportant (that is, to make $\Omega_0$ very
close to unity). We have no argument against that approach, although
in general it seems no easier to arrange potentials for long periods
new inflation in the landscape picture than in other frameworks. The
motivation for this paper is that models with slightly smaller values
of $\Omega_0$ might be considerably more interesting. In fact, any
argument that the generic instance of inflation would tend to be short
(as some have made in the landscape context) would favor a
large effect of the sort we discuss here. 
} 

Our main motivation is the string theory landscape, but we also note
that the puzzling low quadrupole and octopole
($C_2$ and $C_3$) in the WMAP first year data suggest that
interesting information might be lurking in the cosmic perturbations
on large scales\cite{WMAP1,WMAP2}.  Since the open inflation
perturbation spectrum  
depends not only on $\Omega_0$, but on the curvature of the
inflaton potential during the period of old inflation (before
tunneling) in principle we could read information about the physics of
the universe before the big bang from large scale cosmological data. 

With these motivations, we have undertaken a calculation of 
the CMB temperature anisotropies in Open Inflation models.
Unfortunately, our results show that the differences between the open
inflation results and the generic formula used in most cosmology
papers is immeasurably small for realistic cosmological parameters.
Thus we have nothing new to add to the interpretation of
cosmological data.  In particular, at least as far as the Open
Universe models go, the standard determination of the value of
$\Omega_0$ and other cosmological parameters is unaltered, and there is
no opportunity to measure new parameters from other parts of the
string theory landscape. Of course this also means that we cannot rule out open
inflation models with realistic values of $\Omega_0$.  The one caveat
is that our work assumes that the thin wall 
approximation gives a valid treatment of the tunneling event.   It is
possible that corrections to this approximation could lead to a more
interesting result.  

It is also possible that a deeper understanding
of the string theory landscape could lead to other kinds of predictive
power in connection with open inflation.  For example, a ``most
likely'' form for the inflaton driving the post-tunneling period of
new inflation could emerge, which in turn could lead to specific
signature in the CMB power.  This is not the effect we consider in
this paper, which is devoted to effects generic to {\em all} open
inflation models.

\section{The primordial power spectrum}

The primordial power spectrum for Open Inflation presented in \cite{BT2} is 
\begin{eqnarray}
P_{\chi}(\beta)=&&
\frac{9}{4\pi^2}
\left( \frac{H^3}{V,_\phi} \right)^2
\frac{1}{\beta (\beta^2+1)} \nonumber\\
&&\times
\left[\frac{
e^{\pi\beta}+e^{-\pi\beta}
+\frac{|{\cal C}_2|}{{\cal C}_1}
\left(
\frac{\beta+i}{\beta-i}e^{i\bar{\varphi}}
+\frac{\beta-i}{\beta+i}e^{-i\bar{\varphi}}
\right)
}{e^{\pi\beta}-e^{-\pi\beta}}
\right]
\nonumber \\
\end{eqnarray}
where k is the co-moving wavenumber which is related to $\beta$ and
the curvature $K$ by $k^2=\beta^2-K$. 
With the usual normalization, 
$k^2=\beta^2+1$.
The field variable that
gives the density fluctuations is $\chi$, and
${\cal C}_1$ and ${\cal C}_2$ are parameterized by
\begin{equation}
{\cal C}_1=2\pi\cosh^2[\bar{\xi}(\beta)]
\end{equation}
\begin{equation}
{\cal C}_2=2\pi\cosh[\bar{\xi}(\beta)]
\sinh[\bar{\xi}(\beta)]e^{i\bar{\varphi}}
\end{equation}
The definitions of ${\cal C}_1$ and ${\cal C}_2$ are

\begin{equation}
{\cal C}_1=2\pi\left[
1+\frac{\sin^2(\pi\nu')}{\sinh^2(\pi\beta)}
\right]
\end{equation}

\begin{eqnarray}
{\cal C}_2=&&
2\pi\frac
{\sin(\pi\nu')\Gamma(i\beta-\nu')\Gamma(1-i\beta)}
{\sinh^2(\pi\beta)\Gamma(-i\beta-\nu')\Gamma(1+i\beta)}
\nonumber \\
&&\times
\left(
\cosh(\pi\beta)\sin(\pi\nu')-i\sinh(\pi\beta)\cos(\pi\nu')
\right)
\end{eqnarray}

with $\nu'=\sqrt{\frac94-m^2}-\frac12$.  
Here, $m^2$ is the false vacuum effective mass squared
(the second derivative of the potential during the false vacuum inflation) 
in plank mass units. 
It is then more direct to express the power spectrum as

\begin{eqnarray}
P_{\chi}(\beta)=&&
\frac{9}{4\pi^2}
\left( \frac{H^3}{V,_\phi} \right)^2
\frac{1}{\beta (\beta^2+1)}
\nonumber\\
&&\times
\left[
\coth(\pi\beta)+
\frac{(\beta^2-1)Re({\cal C}_2)-2\beta Im({\cal C}_2)}
{{\cal C}_1 (\beta^2+1) \sinh(\pi\beta)}
\right]
\nonumber \\
\end{eqnarray}

In open inflation, rather than having $k^2\chi$ relating to the density
fluctuations, we have $(\beta^2-4K)\chi$ so the primordial power spectrum
with be

\begin{eqnarray}
P(\beta)=&&
\frac{(\beta^2+4)^2}{\beta (\beta^2+1)}
\nonumber \\
&&\times
\left[
\coth(\pi\beta)+
\frac{(\beta^2-1)Re({\cal C}_2)-2\beta Im({\cal C}_2)}
{{\cal C}_1 (\beta^2+1) \sinh(\pi\beta)}
\right]
\nonumber \\
\end{eqnarray}
compared to the standard \cite{ZS1}

\begin{equation}
P(\beta)=
\frac{(\beta^2+4)^2}{\beta (\beta^2+1)}
\end{equation}

Thus, all that is necessary to compare Open Inflation predictions with
the standard results is to insert the bracketed term into the initial
power spectrum in the cmbopen subroutine of CMBfast\cite{CMBFast}.
The bracketed term quickly approaches unity for $\beta > 1$
(wavelengths smaller than the curvature scale), so it effects the very
largest scales with out changing anything on small scales.  For
concreteness we take a tilt of unity ($n_s = 1$).

\section{Evaluation}

\begin{figure}
\scalebox{.4}{\includegraphics{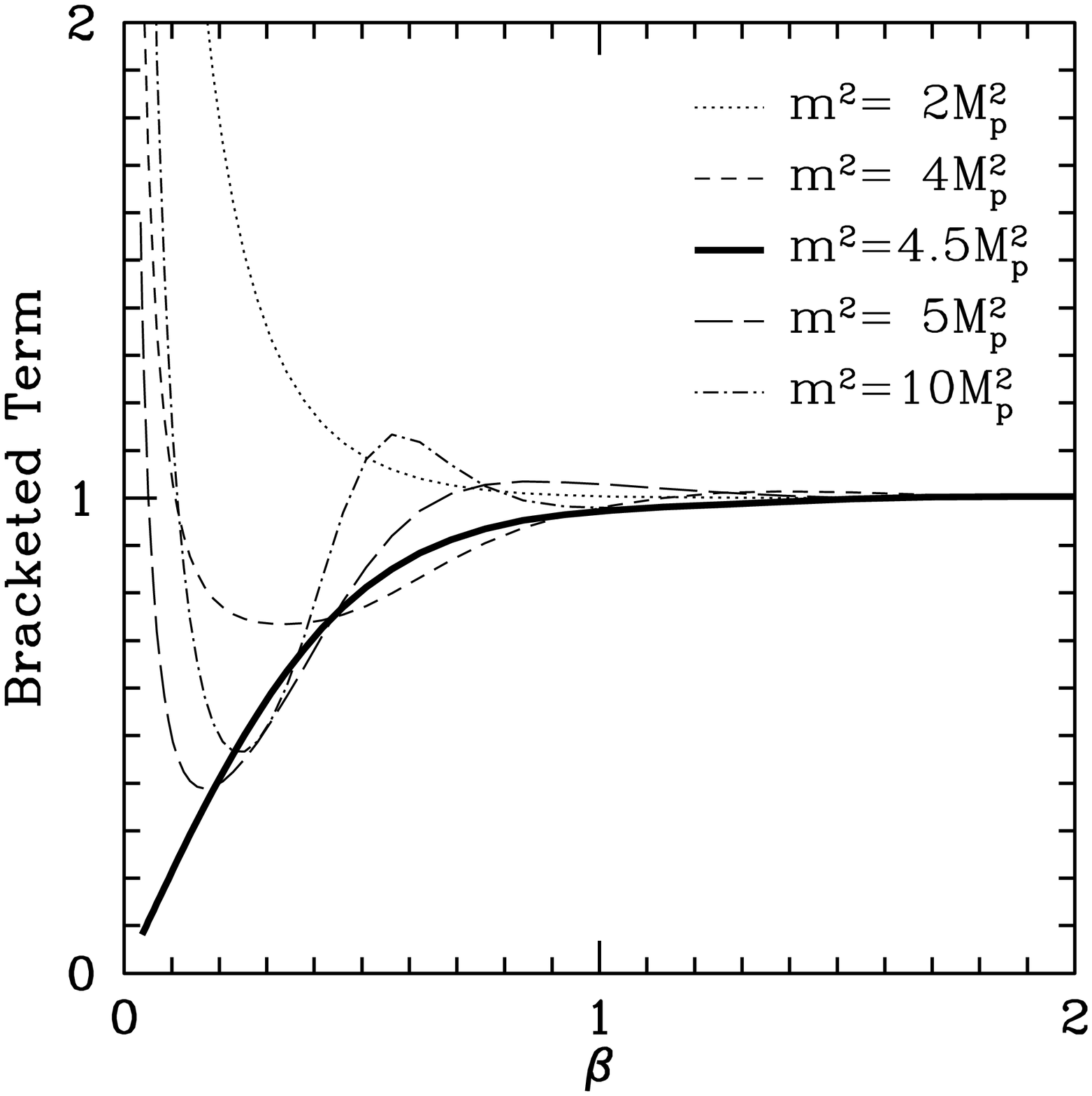}}

\scalebox{.4}{\includegraphics{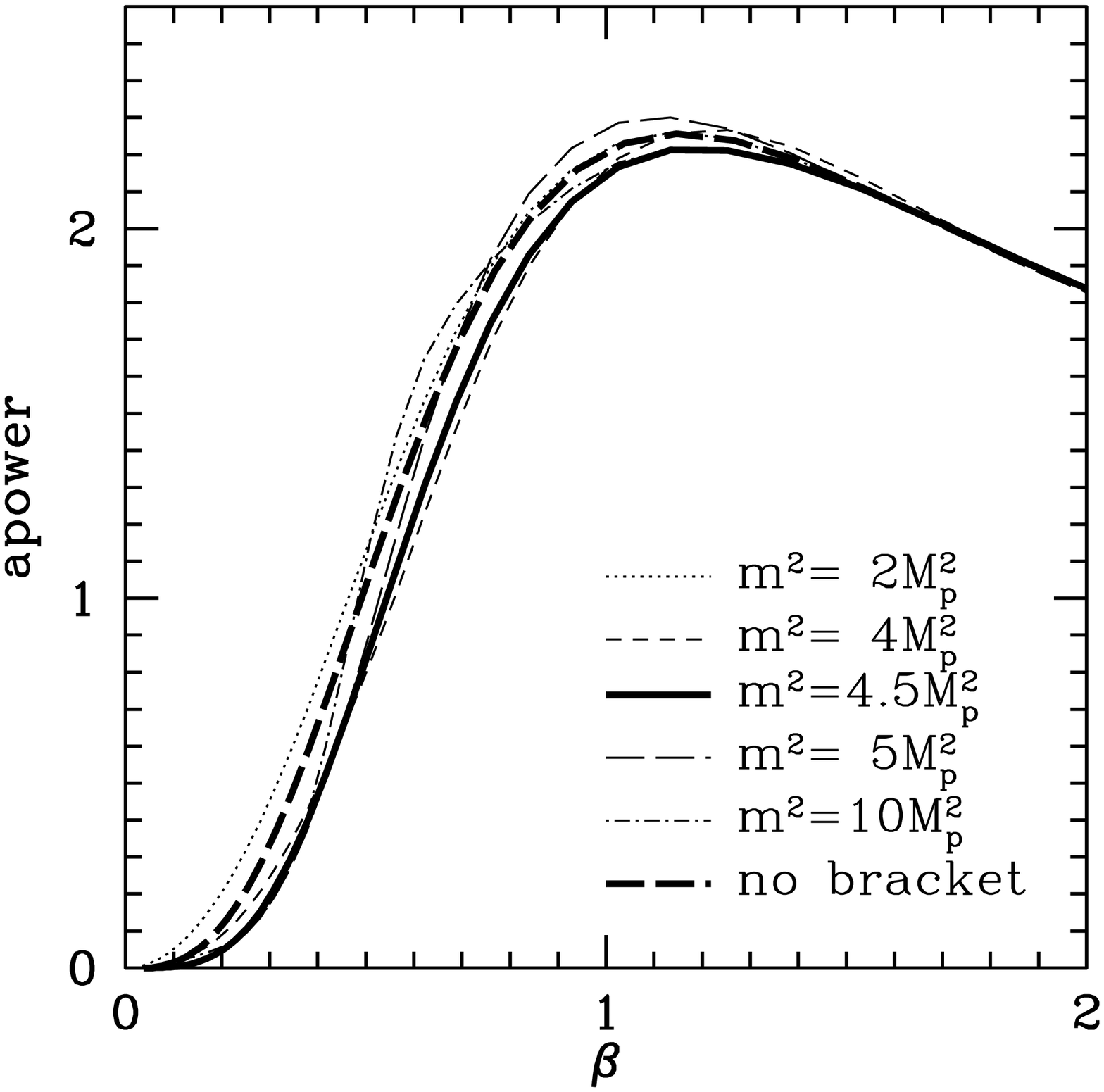}}
\caption{\label{fig1}
Above is a plot of the bracketed term (top) and its effects on the
open universe primordial power spectrum (bottom).
As you can see, the different $m^2$ values give the bracketed term
different oscillations.  The $m^2=4.5$ plot is represented in each
graph as the thick solid line;
the thick dotted line is the power spectrum without the bracketed term.
}
\end{figure}

To evaluate the CMB anisotropies from Open Inflation, the program CMBfast 
\cite{CMBFast} was used to calculate CMB temperature power spectra,
and an expression for the bracketed term was inserted into the subroutine
cmbopen. As illustrated in Fig. \ref{fig1}, we explored the dependence
of the bracketed term on the false vacuum mass and found that it
controlled an oscillation in the bracketed term with respect to
$\beta$. We found that $m^2=4.5 M_p^2$ yielded the strongest
suppression of power at small wave number (of interest because of the
WMAP anomalies). 

An important feature to note is that the open power spectrum tends
toward zero at very small wave numbers without the bracketed term,
leaving only a limited window
of wavenumbers for which the bracketed term has any effect.
The bracketed term does diverge for most choices of $m^2$ as wavenumber
goes to zero, but not fast enough to overcome the rest of the power
spectrum.  
This limits the effect of increasing the curvature on the power spectrum.
Also note that, for $m^2=4.5 M_p^2$, the bracketed term
does not appear to diverge, but rather tends toward zero with no
oscillation.

A wide range of curvatures were tested, comparing power spectra obtained
from identical parameters with and without the correction.
A best fit for parameters with a prior on $\Omega_{tot}$ not being readily
available\footnote{Given the overall disappointing nature of our
results, it was not worth the effort to undertake a full-scale
optimization in a large parameter space}, the choice of the
parameters for these trials is a bit 
arbitrary.
However, the effect of Open Inflation should be independent of all but
the curvature scale, and we are comparing spectra that differ only by
the inclusion or exclusion of the extra term that distinguishes 
Open Inflation.  
Given the small size of
the difference the bracketed term generated, which are summarized in 
Table \ref{table1}, these concerns are largely unimportant.

\begin{table}
\begin{center}
\begin{tabular}{|c|c|}
\hline
$\Omega_{tot}$& \% decrease in $C_2$\\
\hline
.99&$\sim$.01\% \\
.98& .02\% \\
.95& .08\% \\
.90& .20\% \\
.85& .25\% \\
.80& .27\% \\
\hline
\end{tabular}
\end{center}
\caption{\small
This table details the percent decrease in $C_2$,
the $l=2$ value of power,
caused by including the open inflation corrections with
false vacuum mass $m^2=4.5$ in plank units, as this value has
the most effect on the primordial power spectrum.
The first two entries were done using CMBfast with
the best fit parameters given by the WMAP team, 
with the dark energy density reduced to achieve the
stated total density.  The rest were done simply using the CMBfast
default settings with dark energy reduced.  All had no re-ionization. 
We chose $C_2$ to show
here because the effect on the other multipoles was even less significant.
The effect on the $C_l$'s is much less
dramatic than on the quantities shown in the plots because the $C_l$'s
depend on the power at many values of $\beta$, not just at $\beta
\approx 1$ where the effect on the power is most pronounced.
The effect of the correction should depend on curvature and mass alone,
so the values given will at least approximate those for any model with
that curvature.  
} \label{table1}
\end{table}

\section{Conclusions}

The effects of Open Inflation on the CMB power
spectrum are very small compared to the cosmic variance
for the effected observables (Table 1 shows that the most effected
observable, $C_2$, experiences less than a 1\% change, while it has a
cosmic variance $O(50\%)$).  We conclude that the generic form for
perturbations from Open Inflation are not distinguishable in the CMB
temperature anisotropy power spectrum from perturbations given by the standard
formula used throughout the literature.  Thus this data cannot
be used to identify evidence for or against Open Inflation or measure
parameters in other vacuua in the proposed string theory landscape
that might be reflected in the Open Inflation primordial
spectrum. Also, the general differences between the Open Inflation power
spectrum and the standard version are so small that simply  choosing between
the two will not significantly impact constraints on cosmological
parameters from CMB data. However, if our theoretical understanding
evolves to the point where specific inflaton potentials are strongly
preferred, a greater distinguishability  between the two types of
inflation might possibly emerge. 
We note that the power spectrum derived in 
\cite{BT2} uses a thin wall approximation that may not be valid in
many theories, and the effect of relaxing this assumption is unknown.

\section*{Acknowledgments}
%\begin{acknowledgments}
We acknowledge helpful discussions with M. Bucher, N. Kaloper, M. Kaplinghat, 
L. Knox and especially with L. Susskind who stimulated our interest
in this topic. This work was supported in part by DOE grant
DE-FG03-91ER40674.  
%\end{acknowledgments} 

\end{document}